\documentstyle[12pt]{article}
\def\ds{\displaystyle}
\begin{document}
\thispagestyle{myheadings} 
\markright{right head}

\title{A new approach to perturbation theory for a Dirac particle 
in a central field\footnote{To be published in Physics Letters A 260 (1999) 10-16
}}

\author{I.V. Dobrovolska and R.S. Tutik
\footnote{Author to whom all correspondence should be addressed.
E-mail address: tutik@ff.dsu.dp.ua}}

\maketitle

Department of Physics, Dniepropetrovsk State University,
Dniepropetrovsk, UA-320625, Ukraine.
\\  \\
{\sc Abstract.}
The explicit semiclassical treatment of logarithmic perturbation theory 
for the bound-state problem within the framework of the Dirac equation
is developed. Avoiding disadvantages of the standard approach in the 
description of exited states,  new handy recursion formulae with the same
simple form both for ground and exited states have been obtained.
As an example, the perturbation expansions
for the energy eigenvalues for the Yukawa potential containing
the vector part as well as the scalar component are considered.

\newpage

Spectrum analysis poses some of the most important problems  
in quantum mechanics. Success of nonrelativistic potential
models of quark confinement [1] reattracted attention to
the bound-state problem in relativistic physics as well [2,3].
Several attempts have been made to describe relativistic systems in
a central field due to a Lorentz-vector and
Lorentz-scalar interaction within the framework of the Dirac
equation. However, for almost all potentials this equation is not
exactly solvable which compels to resort to some
approximation methods.

 A number of such approaches to solving the radial Dirac equation
in analytical expressions have been developed, including,
in particular, the use of the WKB-method [4,5], the hypervirial
and Hellman-Feynman theorem [6,7],
the 1/$N$-expansion [8-17], the algebraic approach [18], the method
of Regge trajectories [19-21]
and various perturbation schemes [21-26].

Despite such a variety of methods one of the most popular techniques
is still remained logarithmic perturbation theory [27-32].
This technique involves reducing the Dirac equation,
 which becomes a pair of coupled first-order differential equations
in central-field problems, to a nonlinear Riccati equation.
In the case of ground states,
the consequent expansion
in a small parameter results in handy recursion relations which permit
us to derive the high order corrections to energy eigenvalues and
eigenfunctions.
It should be emphasized that
 high orders of expansions are needed for
applying modern summation procedures because
the obtained perturbation series are
typically divergent. However, when radially exited
states are considered, the standard approach becomes extremely
cumbersome and, practically, inapplicable. It is caused by factoring
out zeros of the wave functions, which, in addition, are not the same
zeros for small- and large-components of the Dirac spinor [33].

On the other hand, it is known, that the radial quantum number,
 $n_r$,
most conveniently and naturally is introduced in consideration by
means of quantization conditions, as in the WKB-approach [34,35].
However, since the WKB-approximation is more suitable
for obtaining energy
eigenvalues in the limiting case of large quantum numbers and
the perturbation theory, on the contrary, deals with low-lying
levels, the WKB quantization conditions need change.

Recently, a new technique based on a specific
quantization condition has been proposed to get
the perturbation series via semiclassical expansions
within the one-dimensional Schr\"{o}dinger
equation [36]. For the Dirac equation,
on performing the scale transformation,
$r\to\hbar^{2}r$, the coupling constants appear
in common with powers of Planck's constant, $\hbar$,
thus implying the possibility to obtain perturbation expansions in
a semiclassical manner in this case, too.

The objective of this letter is to develop the explicit semiclassical
treatment of logarithmic perturbation theory for the bound-state
problem within the framework of the Dirac equation and to describe
a new procedure for deriving perturbation corrections through handy
recursion formulae having the same simple form both for ground
and exited states.

The proposed technique can be regarded as a further investigation
of a part assigned to a rule of achieving a classical limit in construction
of semiclassical methods. In addition to the rule,
$ \hbar\to 0,\;n_r \to\infty,\; l\to\infty,\;\hbar
n_r={\rm const},\;\hbar l={\rm const},$
required within the WKB-approach; and conditions,
$ \hbar\to 0,\; n_r={\rm const},\;
l\to\infty,\;\hbar n_r\to 0,\;\hbar l={\rm const}$,
which are applied within the method of 1/N-expansion [17];
here we address ourselves  to the alternative possibility:
$ \hbar\to 0,\; n_r={\rm const},\;l={\rm const},\;
\hbar n_r\to 0,\;\hbar l\to 0$, that results
in the explicit semiclassical treatment of the logarithmic
perturbation theory.
  \\

(1) {\it Method. }
We study the bound state problem for a
single fermion moving in an attractive central potential.
This potential contains both the time component of a Lorentz
four-vector, $ V(r)$, and a Lorentz-scalar term, $W(r)$,
which, in general, have a Coulomb-like behaviour at the origin
\begin{equation}
V(r) =\frac{1}{r}{ \sum_{i=0}^{\infty}{V_i}\,r^i}\;,\\
W(r)=\frac{1}{r}{ \sum_{i=0}^{\infty}W_i r^i\; },
 \end{equation}
though the case $ V_0=0$ or $ W_0=0 $ is permissible, too.
In what follows, a scalar potential will be included in the mass
term $ m(r) $ by analogy with "dynamical mass" models of quark
confinement [2,3]
\begin{eqnarray}
 m(r)=m+ \frac{ W(r) }{ c^2 }  \; .
 \end{eqnarray}
Then the Dirac radial wave equations have the form
\begin{eqnarray}
\hbar F'(r)-\frac{\hbar\chi}{r}F(r)+\frac{1}{c}[{E-V(r)-m(r)c^2}]G(r)=0\;,   \nonumber \\
\hbar G'(r)+\frac{\hbar\chi}{r}G(r)-\frac{1}{c}[{E-V(r)+m(r)c^2}]F(r)=0\;.
\end{eqnarray}
Here $ F(r) $ and $ G(r) $ are the small and large components of the
 wavefunction of a particle and $ E $ is its total energy,
$\chi =s(j+{ \frac{1}{2} }) $ for $ j=l-{\frac{s}{2} } $, with $ s=\pm1 $
denoted the sign of $\chi $.

Eliminating $ F(r) $ from the system (3) and performing the substitution,
$ R(r)= \hbar G'(r) / G(r) $, for the logarithmic derivative of large
component we then arrive at the Riccati equation
\begin{equation}
\begin{array}{l}
\hbar R'(r)-\hbar Q(r) R(r)+R^2(r) \\
\ds =\frac{\hbar^2 \chi(\chi+1)}{r^2}
+Q(r)\frac{\hbar^2\chi}{r}+m^2(r)c^2-\frac{1}{c^2}[E-V(r)]^2\;,
\end{array}
\end{equation}
with $Q(r)=[m'(r)-V'(r)]/[E+m(r)-V(r)]$.

As was above pointed out, logarithmic perturbation theory does involve
coupling constants in common with powers of Planck's constant.
Therefore we attempt now to solve eq.(4) in a semiclassical manner.
Taking into account the leading orders in $\hbar$ of the quantities 
$E\sim  \frac{1}{\hbar^2} $, $\hbar \chi \sim \hbar $, from Riccati
equation (4) we have
 \begin{equation}
\begin{array}{l}
\ds R(r)= \frac{1}{\hbar}  \sum_{i=0}^{\infty}R_i(r) \hbar^{2i}\;, \\
\ds Q(r)=\hbar^2 \sum_{i=0}^{\infty} Q_i(r)\hbar^{2i} \;, \\
\ds E= \frac{1}{\hbar^2}  \sum_{i=0}^{\infty}E_i\hbar^{2i}\, .
\end{array}
\end{equation}

Keeping in mind that relativistic mechanics must go over to nonrelativistic
one as the speed of light tends to infinity
and quantum mechanics must go over
to classical one as $\hbar\to 0$, the correlation between
these constants
has to be set. By analogy with quantum electrodynamics the
foregoing consideration will be carried out under the condition
$\hbar c\sim O(1)$ [37]. Moreover, for simplicity we put
$\hbar c=1 $.

On substituting the expansions (5) into the Riccati equation (4)
and comparing coefficients of the different powers of $\hbar$,
one obtains the following hierarchy of equations
\begin{equation}
\begin{array}{l}
\ds R_0^2(r)=m^2-E_0^2\;, \\
\ds R_0(r)R_1(r)=E_0\Bigl[  V(r)-E_1 \Bigr]  +mW(r) \;, \\
\ds R_1'(r)+2R_0(r)R_2(r)+ R_1^2(r) - R_0(r) Q_0(r) \\
\ds =\frac {\chi(\chi+1)}{r^2}+
2E_1V(r)-E_1^2-2E_0E_2+W^2(r)-V^2(r)\;, \\
\cdots  \\
\ds R_{k-1}'(r)+\sum_{j=0}^kR_j(r)R_{k-j}(r)
-\sum_{j=0}^{k-2}R_j(r)Q_{k-j-2}(r) \\
\ds =2E_{k-1}V(r)-\sum_{j=0}^kE_jE_{k-j}
+\frac{\chi}{r}Q_{k-3}(r) \;, \\
\end{array}
\end{equation}
where

 $
\ds Q_0(r)=\frac{1}{E_0+m}\Bigl[ W'(r)-V'(r)\Bigr] , $

$ \ds Q_k(r)=-\frac{1}{E_0+m}\left[\sum_{j=0}^{k-1}
{Q_j(r)E_{k-j}}+Q_{k-1}(r)\Bigl[W(r)-V(r)\Bigr]\right].
$

For nodeless states this system can be solved straightforwardly.
 However, when radial excitations are described with standard
technique the nodes of the wave functions need to be factored
out first and consideration becomes extremely cumbersome.
We intend to circumvent this difficulty by making use of the
quantization condition. Its fundamental idea that stems from the
 WKB-approach [34,35] is well known as the principle of argument in
the analysis of complex variables. Being applied to the logarithmic
derivative, $R(r)$, it means that
\begin{equation}
\frac{1}{2\pi\rm{i}}\oint{R(r)\,{\rm d} r}=\hbar N\;,
\end{equation}
where $N$ is a number of zeros inside a closed contour.

This condition is exact and is widely used for deriving the high-order
corrections to the WKB-approximation and the 1/$N$-expansions.
There is, however, one important point to note. The radial and orbital
quantum numbers, $n_r$ and $l$, correspondingly, are specific quantum
 notions and need be defined before going over from quantum
mechanics to classical physics. Therefore the quantization condition (7)
 must be supplemented with the rule of achieving a classical limit that
stipulates the type of semiclassical approximation.

In particular, within the WKB-approach the passage to the classical
limit is implemented using the rule
 \begin{equation}
  \hbar\to 0,\;n_r\to\infty,\; l\to\infty,
\;\hbar n_r={\rm const},\;\hbar l={\rm const},
\end{equation}
whereas the 1/$N$-expansion, being complementary to the
 WKB-method, requires the conditions [17]
\begin{equation}
  \hbar\to 0,\; n_r={\rm const},\;l\to\infty,
\;\hbar n_r\to 0,\;\hbar l={\rm const}.
\end{equation}

The semiclassical treatment of logarithmic perturbation theory proved
to ivolve the alternative possibility:
\begin{equation}
  \hbar\to 0,\; n_r={\rm const},\;l={\rm const},
\;\hbar n_r\to 0,\;\hbar l\to 0.
\end{equation}

Notice that the part of this rule concerned the orbital quantum number,
 $l$, differs from one used within the WKB-approach and the 1/$N$-expansion
method. In our consideration this part, implying the first order in $\hbar$ for
the quantity $\hbar\chi$, has been used in deriving the system (6).

The remaining part of the rule respects the radial quantum number.
Due to (10) the right-hand side of
 the equality (7) has the first order in $\hbar$ and, hence,
 on substituting the
 expansion (5) the quantization conditions (7) takes the form
\begin{equation}
  \frac{1}{2\pi\rm{i}}\oint{R_i(r)\,{\rm d} r}=N\delta_{i,1} \; ,
\end{equation}
where the Kronecker delta $\delta_{ij}$ is used.

Before proceeding further, we must specify the quantity $N$. In contrast
to the WKB-approach and the 1/$N$-method, we choose such a contour of
integration which encloses both the nodes of the wave function $G(r)$
and the boundary point, $r=0$.
Then the quantity N is depends on both the radial quantum number,
$n_r$, and the behaviour of the wave function near the origin, and
is given by

\begin{equation}
  N=n_r+\sqrt{\chi^2+W_0^2-V_0^2}\;\; ,\;n_r=n+(s+1)/2\;\;,\;n=0,1,2,...
\end{equation}

A further application of the theorem of residues to the explicit
form of functions $R_i(r)$ easily solves the problem of taking into
account nodes of the wave functions for exited states.
 \\

(2) {\it Recursion formulae}. We begin with investigation
of behaviour of the functions $R_i(r)$. From the system (6) we have
\begin{equation}
  R_0(r)=-\sqrt{m^2-E_0^2}\;,
\end{equation}
where the minus sign is chosen from boundary condition.
Then the function $R_1(r)$ has a simple pole at the origin,
owing to the Coulombic behaviour of the potentials
at this point, while the function $R_k(r)$ has a pole of the
order $k$. Hence $R_k(r)$ can be represented by the Laurent series
\begin{equation}
  R_k(r)=r^{-k}\sum_{i=0}^\infty{R^{k}_{i}r^i}\;,
\end{equation}
which makes it possible to write the quantization conditions (11) as
\begin{equation}
  R_k^{k+1}=N
\delta_{k,0}\;.
\end{equation}

Now, by analogy with $R_k(r)$, let us also represent the functions
$Q_k(r)$, involved in the expansion (5), as a power series in $r$:
\begin{equation}
  Q_k(r)=r^{-2-k}\sum_{i=0}^\infty{Q^{k}_{i}r^i}\;.
\end{equation}

From the equation $Q(r)[E+m(r)-V(r)]= m'(r)-V'(r) $ we then obtain
\begin{equation}
\begin{array}{l}
  \ds Q_i^0=\frac{i-1}{E_0+m}(W^0_i-V^0_i), \\
\ds Q_i^k=- \frac{1}{E_0+m}\left[ \sum_{j=0}^{k-1}{Q^j_{j+i-k}E_{k-j}}
+\sum_{j=0}^{i}{Q^{k-1}_j(W_{i-j}-V_{i-j})} \right].
\end{array}
\end{equation}

Finally, substituting expansions (14) and (16) into the last
equation of the system (6) and collecting coefficients of the
like powers of $r$ leads to the recursion relation in terms of the
Laurent coefficients, $R^k_i$:
\begin{equation}
\begin{array}{l}
\ds  R^k_i = -\frac{1}{2R^0_0}
\biggl[
     (i-k+1)R^{k-1}_i+
     \sum^{k-1}_{j=1}  \sum^i_{p=0} R^j_p R^{k-j}_{i-p}
     -\sum^{k-2}_{j=0} \sum^i_{p=0} Q^j_p R^{k-2-j}_{i-p}
\\
  \ds   -\chi Q^{k-3}_i
     +\delta_{k,i} \sum^k_{j=0} E_j  E_{k-j} -2 E_{k-1} V_{i-k+1}
\\
  \ds    +\delta_{k,2}  \sum^i_{p=0} (V_p  V_{i-p}- W_p  W_{i-p})
       -\delta_{k,1} \, 2 m W_i
       -\delta_{i,0} \delta_{k,2}\,\chi ( \chi + 1 )
\biggr]\; ,
\end{array}
\end{equation}
where for universality of designations we put $R_0^0=R_0,R^0_i=0,i>0.$

In the case $i\not=k$, this formula is intended for obtaining
coefficients $R_i^k$. When $i=k$, by equating the explicit
expression for $R^{k+1}_k$ to the quantization condition (15)
we get the recursion relation for the energy eigenvalues
\begin{equation}
\begin{array}{l}
 \ds E_0 = \frac{ m}{N^2+V_0^2}
\left(  N   \sqrt{N^2 + V_0^2 - W_0^2 } - V_0  W_0   \right)
 \;,k=0,
\\ \\
\ds E_k=\frac{R^0_0R^1_0}{2(E_0R^1_0+V_0R^0_0 )}
\Biggl[
      \frac{1} {R^1_0}
      \Biggl(
             \sum^{k-1}_{j=2}\sum^k_{p=0} R^j_p R^{k+1-j}_{k-p}
               +2   \Theta   ( k-2 )  \sum_{p=1}^k R^1_p R^{k}_{k-p}
         \\
\ds
               -\sum^{k-1}_{j=0} \sum^k_{p=0} Q^j_p  R^{k-1-j}_{k-p}

 +\delta_{k,1}  (V_0V_1- W_0W_1)
        -  \chi Q^{k-2}_k
   \Biggr)
\\
\ds -\frac{1}{R_0^0}
     \Biggl(
       R^{k-1}_k
       +\sum^{k-1}_{j=1}\sum^k_{p=0}R^j_p  R^{k-j}_{k-p}
        -\sum^{k-2}_{j=0}\sum^k_{p=0}Q^j_pR^{k-2-j}_{k-p}
\\ \ds
        +\sum^{k-1}_{j=1} E_j E_{k-j}
  -2E_{k-1} V_{1}
    +\delta_{k,2}\sum^2_{p=0}(V_p  V_{2-p}- W_pW_{2-p})
\\ \ds
     -\delta_{k,1} 2 m W_1
     - \chi Q^{k-3}_k
\Biggr)
\Biggr] \;,k>0\;.
\end{array}
\end{equation}
Here $E_0$ does be the exact solution to the Dirac-Coulomb
equation [38,39] and we use the step function

$\begin{array}{ll}
\Theta(k)&=1,\quad k\geq0,\\
              &=0,\quad k<0\; .
\end{array}
$

Thus, equations (18) and (19) determine the coefficients
of perturbation expansions of energy eigenvalues and
eigenfunctions for screened Coulomb potentials in the
same form both for ground and exited states.
\\

%
%
(3) {\it Examples of application.}
As a check of the obtained formulae we calculate the energy
eigenvalues for the pure-vector, screened Coulomb potential
of general form. On applying the recursion relations (18)
and (19), analytical expressions for the perturbation
coefficients are found to be equal to
\begin{equation}
\begin{array}{rl}
E_0= & \ds \frac{m N}{\sqrt{N^2+a^2}} \; ,
\\
E_1=&a\,{V_1}\; ,
\\
E_2=&\ds -\frac{V_2}{2\,\rho^2}\left(3a^2\epsilon-\chi(\chi \epsilon+1 )\rho^2
 \right)\; ,
\\
E_3=&\ds \frac{V_3}{2\,\rho^4}\left(
a^3(4 \epsilon^2+1) -a(2\chi^2\epsilon^2+3\chi \epsilon+\chi^2-1) \rho^2
\right) \; ,
\\
E_4=&\ds\frac{1}{8\,\rho^6}
\Bigl( V_2^2
        \Bigl[
                 a^4\,\epsilon ( 5 \epsilon^2-12 )  +
         a^2 \epsilon ( 6\chi^2 \rho^2-5)\rho^2  {}
  \\ &  \ds
                +  \chi^2 \bigl[
                \chi^2\epsilon ( \epsilon^2+2 ) +
                 \chi(  4 \epsilon^2+2 )  + 3\epsilon
                         \bigr] \rho^4
        \Bigr]  {}
 \\  &   \ds
       + V_4
        \Bigl[
                  -   5 a^4 \epsilon ( 4 \epsilon^2+3 )
                  +  a^2 \bigl[
                              6 \chi^2 \epsilon ( 2 \epsilon^2+3)
                            +  6 \chi  ( 4 \epsilon^2+1)
                             - 25\epsilon
                            \bigr] \rho^2  {}
 \\ &  \ds
             -3  \chi ( \chi^2-1)
                     ( \chi \epsilon+2) \rho^4
       \Bigr]
\Bigr) \;,
\\
E_5=&\ds\frac{-a}{8\,\rho^8}
\Bigl( V_2 V_3
             \Bigl[ 3 a^4
            ( 8\epsilon^4-20 \epsilon^2 -3)  -
           a^2 \bigl[   \chi ^2( 32\epsilon^4  - 36\epsilon^2 -10 )

 \\ &   \ds   +  \chi \epsilon  (  10\epsilon^2 -24)
             + 9(  6\epsilon^2 +1 )
                \bigr]   \rho^2
                            + \chi \bigl[ 30\chi ^2\epsilon^3

\\ &  \ds
                            + 24\chi \epsilon^2 + 10\epsilon
                            +\chi
                            + \chi ^3( 8\epsilon^4  + 8\epsilon^2 -1)
                    \bigr]   \rho^4
           \Bigr]
\\ &  \ds
    + V_5   \Bigl[
                        - 3a^4 (  8\epsilon^4 + 12\epsilon^2 +1 )
        + a^2 \bigl[    \chi ^2(  16\epsilon^4+ 48\epsilon^2 + 6 )
      \\ &  \ds
                         + 10\chi \epsilon(  4\epsilon^2+3 )

                        -  15( 6\epsilon^2+1 )
                \bigr] \rho^2
\\ &   \ds
                + \bigl[ 5\chi ^2(  4\epsilon^2+3 )
                 -  3\chi ^4(  4\epsilon^2+1)
                                     +50\chi \epsilon
                              - 30\chi ^3\epsilon-12

                           \bigr]  \rho^4
                    \Bigl]
      \Bigr)\; ,
\end{array}
\end{equation}
where  $ a=V_0 $, $\epsilon =E_0$ and $ \rho $ is $ \sqrt{1-\epsilon^2} $.

One can verifies that the first three corrections coincide with those
derived by McEnnan et al [24] with standard technique.

The next example will be the attractive Yukawa potential, often
utilized in relativistic calculations, which has not only the
Lorentz-vector component, $ V(r)=-(a/r) \, e^{-\lambda r}$, but the
Lorentz-scalar term, $ W(r)=-(b/r) \, e^{-\mu r}$, as well. Now the
analytic expressions for perturbation corrections to the bound state
energy take the form
\begin{equation}
\begin{array}{rl}
  E_0 =& \ds\frac{ m  }
{N^2+a^2}\left(  N   \sqrt{N^2 + a^2 - b^2 } - a b   \right)\;,
\\  \\
E_1=&\ds a\,\lambda  + b\,\mu \,{\epsilon}\;,
\\ \\
E_2=&-\ds \frac{1}{4( b\epsilon + a ) \rho^2}
\biggl(
\lambda ^2 \Bigl[ 3a^3\epsilon - a\chi (  \chi \epsilon+1 ) \rho^2
         + 2a^2b ( 2\epsilon^2+1 )
\\& \ds
  + a b^2\epsilon(  \epsilon^2 +2 )
                  \Bigr]
\\& \ds

 +\mu ^2 \Bigl[ 3b^3\epsilon^2
   - b\chi \epsilon\rho^2 - b\chi ^2\rho^2
         + 2b^2a\epsilon(  \epsilon^2 +2)  +
          ba^2( 2\epsilon^2+1)  \Bigr]
      \biggr)\;,
\\
E_3=&\ds \frac{1}{12( a + b\epsilon ) \rho^4}
\biggl(
 \lambda ^3  \Bigl[
a^4( 4\epsilon^2+1) - a^2(
           2\chi ^2\epsilon^2 + 3\chi \epsilon+ \chi ^2 -1   ) \rho^2

  \\&+  3a^3b\epsilon( 2\epsilon^2+3 )
      +  a^2b^2( 2\epsilon^4+   11\epsilon^2 +2 )

-  ab( 3\chi  +
           (3\chi ^2  -1  ) \epsilon ) \rho^2
\\&
       + ab^3\epsilon(  3\epsilon^2 + 2)   \Bigr]
      \\&

       +   \mu ^3 \Bigl[ b^4(  - 8\epsilon^4 +13\epsilon^2)
       +  b^2(  3\chi \epsilon^3+( 2\chi ^2 + 1 ) \epsilon^2
        - 6\chi \epsilon -5\chi ^2
           ) \rho^2
\\&
  -  3b^3a\epsilon(  2\epsilon^4-\epsilon^2 -6 )
        -b^2a^2(   4\epsilon^4 - 14\epsilon^2 -5)
\\&

 - ba\epsilon(  3\chi \epsilon+ 3\chi ^2-1
           ) \rho^2

 +ba^3\epsilon (2\epsilon^2+3)   \Bigr]
         \\&
       +  \lambda ^2\mu  \Bigl[ 9a^3b\epsilon \rho^2

    -   6a^2b^2(  2\epsilon^4 - \epsilon^2-1 )
\\&
       + 3ab^3\epsilon\rho^2 (  \epsilon ^2 +2 )  -
        3ab\chi (  \chi \epsilon +1 ) \rho^4  \Bigr]
\biggr).
     \end{array}
\end{equation}

In order to assess the speed and accuracy of the perturbation
technique for the Yukawa potential with various ratios of its
component we consider energy eigenvalues for the pure vector
case, $V(r)=- (a/r)\, e^{-\lambda r}$; the pure scalar
potential, $W(r)=-( a/r)\, e^{-\lambda r}$; and the equally mixed
interaction, $V(r)+W(r)=-(I+\gamma_0)(a/2r)\, e^{-\lambda r}$.
Typical results of calculation are represented in Table 1 where
the sequence of the sums of first terms from our expansion
for relativistic binding energies
is compared with the results, $E_{num}$ (in KeV), obtained by
numerical integration. The calculation has been performed for
$s=1,\,n_r=1,\,l=1$ and $s=-1,\,n_r=1,\,l=0$ states with parameters
$ a= \alpha z ,\; \lambda =1,13 \alpha z^{1/3},\; z=74 $ ( $ \alpha $  is the
fine-structure constant and $z$ the nuclear charge).

As it can be seen from Table 1, in all cases we have two
subsequences bounded below and above
the energy eigenvalues. The average of these subsequences at the point of
their maximal drawing together is proved to
result in a quite good approximation to the exact value.
\vskip 1cm

To summarize, we have developed a semiclassical treatment
of logarithmic perturbation theory for a Dirac particle
in a central field. Based upon the $\hbar $-expansions and
suitable quantization conditions, new handy recursion relations
for solving the bound-state problem for the Dirac equation
with the screened Coulomb potential having both vector and
scalar component have been derived. Avoiding the disadvantages
of the standard approach these formulae have the same simple
 form both for ground and exited states and provide,
in principle, the calculation of the perturbation corrections
up to an arbitrary order in the analytic or numerical form.
And at last, this approach does not imply knowledge of
the exact solution for zero approximation, which is obtained
automatically.
$$ $$
This work was supported in part by the International Soros
Science Education Program (ISSEP) under grant APU052102.

\eject

\eject
Table 1
\bigskip

\begin{tabular}{c|lll|lll}
\hline
\vphantom{\Big(}
k & \multispan{2}{\hss $s=1, n_r=1, l=1$} &
& \multispan{2}{\hss $s=-1, n_r=1, l=0$}& \\
\cline{2-7}
\vphantom{\Big(}   & $E_V$ & $E_W$ & $E_{V+W}$
  & $E_V$ & $E_W$ & $E_{V+W}$ \\
\hline
 0&     20.644616&   16.59173&   18.292804&   20.644616&   16.59173&     18.292804\\
 1&     11.091653&   7.348948&   10.846326&   11.091653&   7.348948&     10.846326\\
 2&     12.415123&   9.066156&   11.810855&   12.721342&   9.372375&     12.003761\\
 3&     12.264120&   8.784152&   11.703753&   12.500951&   8.995036&     11.855249\\
 4&     12.308677&   8.880691&   11.730813&   12.558797&   9.118034&     11.890052\\
 5&     12.292914&   8.838096&   11.722213&   12.537837&   9.063240&     11.878837\\
 6&     12.299805&   8.860978&   11.725555&   12.546910&   9.092488&     11.883163\\
 7&     12.296473&   8.847251&   11.724110&   12.542466&   9.074820&     11.881275\\
 8&     12.298236&   8.856234&   11.724792&   12.544844&   9.086483&     11.882175\\
 9&     12.297242&   8.849965&   11.724449&   12.543482&   9.078246&     11.881715\\
10&     12.297834&   8.854576&   11.724631&   12.544306&   9.084385&     11.881963\\
11&     12.297465&   8.851033&   11.724530&   12.543784&   9.079601&     11.881823\\
12&     12.297704&   8.853860&   11.724588&   12.544128&   9.083474&     11.881905\\
13&     12.297544&   8.851529&   11.724553&   12.543893&   9.080233&     11.881856\\
14&     12.297654&   8.853509&   11.724575&   12.544058&   9.083025&     11.881886\\
15&     12.297576&   8.851782&   11.724561&   12.543939&   9.080555&     11.881867\\
\hline
\vphantom{\Big(}
$E_{\rm num}$&     12.297609&   8.852592&   11.724567&   12.543990&   9.081723&     11.881875\\
\hline
\end{tabular}

\end{document}